\documentclass[twocolumn,aps,prb,showpacs,preprintnumbers,amsmath,amssymb,psfrac]{revtex4}

\usepackage{epsfig}
\usepackage{graphicx}
\usepackage{dcolumn}
\usepackage{bm}

\renewcommand{\dag}{^{\dagger}}

\begin{document}
%

\title{
One-dimensional conductance through an arbitrary potential 
}
\author{Tobias Stauber}
\affiliation{Institut f\"ur Theoretische Physik, Ruprecht-Karls-Universit\"at Heidelberg, Philosophenweg 19, D-69120 Heidelberg, Germany
}
\begin{abstract}
The finite-size Tomonaga-Luttinger Hamiltonian with an arbitrary potential is mapped onto a non-interacting Fermi gas with renormalized potential. This is done by means of flow equations for Hamiltonians and is valid for small electron-electron interaction. This method also yields an alternative bosonization formula for the transformed field operator which makes no use of Klein factors. The two-terminal conductance can then be evaluated using the Landauer formula. We obtain similar results for infinite systems at finite temperature by identifying the flow parameter with the inverse squared temperature in the asymptotic regime.  
\end{abstract}
%
\pacs{
71.10.Pm, 73.22.-f, 73.63.-b, 05.10.Cc}
%
%
%
\maketitle
The conductance of one-dimensional (1D) conductors at low temperatures is determined by a sophisticated interplay between impurity scattering and electron-electron interaction. The question was first raised about thirty years ago by Luther and Peschel\cite{Lut74} and Mattis\cite{Mat74} in the context of enhanced conductivity, present in 1D structures at low temperatures. Their results were later confirmed and extended  using renormalization group techniques.\cite{Ape82,Kan92} Recently, evidence was found that the conductance in single-walled carbon nanotubes\cite{Boc99} and GaAs/AlGaAs quantum wires\cite{Rot00} behaves according to the so-called Luttinger liquid theory.\cite{Hal81} 

The above theoretical descriptions were only valid for large systems at low temperatures due to the perturbative treatment in the impurity strength. This shortcoming was overcome by Monte-Carlo calculations\cite{Moo93} and applying the thermodynamic Bethe ansatz.\cite{Fen95} Nevertheless, the latter method already fails for a double impurity. 

To describe general external potentials, it thus seems favorable to tackle the problem from the other end, i.e., to treat the impurity exactly but include the electron-electron interaction perturbatively. Since the interaction of 1D systems is marginal in the renormalization group sense,\cite{Sol79} this has to be done with care. One approach is to use scattering theory within the Born approximation and renormalization group techniques.\cite{Yue94,Sen02} The interaction can also be included via exact renormalization group equations based on the one-particle irreducible effective action.\cite{Med02} 

Recently, we have employed a flow equation scheme for Hamiltonians\cite{Weg94} which is applicable for small electron-electron interaction but arbitrary impurity strength.\cite{Sta03} For infinite systems at zero temperature, the approach yielded a logarithmic expansion for the spectral weight at the impurity site involving the exact boundary exponent. In this Brief Report, we will extend the analysis and recover the exact power-law behavior. For this, we will derive the exact Green's function of the Tomonaga-Luttinger model without making use of the bosonization formula. We will also demonstrate how the conductance can be obtained as a function of the system size and the temperature, respectively, using the Landauer formula.\cite{Lan70}  

We start our discussion from the spinless Tomonaga-Luttinger Hamiltonian\cite{Tom50} of finite length $L$
\begin{align}
H_{TL}\equiv H_0+H_{ee}\label{TLmodel}
\end{align}
with
\begin{align}
H_0=\sum_q\omega_q^0b_q\dag b_q\;,\;
H_{ee}=\frac{1}{2}\sum_qu_q^0(b_qb_{-q}+b_{-q}\dag b_q\dag)\quad.\notag
\end{align}
The initial conditions are given by $\omega_q^0=v_F|q|(1+v_q/2\pi v_F)$ and $u_q^0=|q|v_q/2\pi$ with $q=\pm 2\pi n_q/L$, $n_q\in\mathbb{N}$ where $v_F$ denotes the Fermi velocity and $v_q=v_{-q}$ is the Fourier transform of the electron-electron interaction. The sum is over all non-zero wave numbers $q\neq0$ and the normalized operators of the electron density fluctuations $b_q^{(\dagger)}$ obey canonical commutation relations. They are given by $b_{\pm|q|}=\frac{1}{\sqrt{n_q}}\sum_k c_{k,\pm}\dag c_{k\pm |q|,\pm}$ where $c_{k,\pm}^{(\dagger)} $ are the fermionic annihilation (creation) operators of the plane wave with wave number $k$ of the right- and left-movers and $k=2\pi n/L$ is unbounded ($n\in\mathbb{Z}$). 

We now perform a continuous unitary transformation, $U(\ell)$, such that the interaction Hamiltonian $H_{ee}$ tends to zero for $\ell\to\infty$. In differential form, this transformation can be characterized by the so-called flow equations $\partial_\ell H(\ell)=[\eta(\ell),H(\ell)]$ with the anti-Hermitian generator $\eta(\ell)=-\eta\dag(\ell)$.\cite{Weg94} The parameters of the Hamiltonian (\ref{TLmodel}) thus become functions of the flow parameter $\ell$. This will be indicated by dropping  the upper zero which was introduced in order to specify the initial conditions. 

In this work, we choose the generator as $\eta=[H_0,H_{ee}]$ even in the presence of an external potential.\cite{FootG} This yields
\begin{align}
\eta(\ell)=\frac{1}{2}\sum_q\eta_q(\ell)(b_qb_{-q}-b_{-q}\dag b_q\dag)
\end{align}
with $\eta_q(\ell)=-2\omega_q(\ell)u_q(\ell)=-\text{sgn}(v_q)\tilde\omega_q^2[\sinh(4\tilde\omega_q^2\ell+C_q)]^{-1}$ where $\sinh(C_q)=\tilde\omega_q^2/2\omega_q^0|u_q^0|$ and $\tilde\omega_q\equiv\omega_q(\ell=\infty)$. The fixed point dispersion $\tilde\omega_q$ follows from the flow invariant $\omega_q^2(\ell)-u_q^2(\ell)=v_F^2q^2(1+v_q/\pi v_F)$ as $u_q(\ell)\to0$ for $\ell\to\infty$. For attractive electron-electron interaction we impose the lower bound $v_q>-\pi v_F$ in order to avoid phase separation.\cite{Mat65}

The $\ell$-dependent parameters then read
\begin{align}
\label{TLSolution}
\omega_q(\ell)&=c_q(\ell)\omega_q^0+s_q(\ell)u_q^0\;,\;
u_q(\ell)=c_q(\ell)u_q^0+s_q(\ell)\omega_q^0
\end{align}
where $c_q(\ell)\equiv\cosh(E_q(\ell))$ and $s_q(\ell)\equiv\sinh(E_q(\ell))$ with 
\begin{align} 
\label{Connection}
E_q(\ell)=\int_0^\ell d\ell'\eta_q(\ell')=\ln\Big(\frac{\tanh(2\tilde\omega_q^2\ell)K_q^2+1}{\tanh(2\tilde\omega_q^2\ell)K_q^{-2}+1}\Big)^{\frac{1}{4}}
\end{align}
and the Luttinger liquid parameter $K_q^2=(\omega_q^0-u_q^0)/(\omega_q^0+u_q^0)$. 

In order to evaluate correlation functions, the observable has to be transformed by the same continuous unitary transformation, i.e., $\partial_\ell\psi_\pm(x,\ell)=[\eta(\ell),\psi_\pm(x,\ell)]$.
Starting from the fermionic representation of the field operator, i.e., $\psi_\pm(x)=L^{-1/2}\sum_ke^{ikx}c_{k,\pm}$, the flow equations generate an infinite hierarchy of operators. Nevertheless, this expansion factorizes and  only involves one $q$-dependent function of $\ell$, $\varphi_q^\pm(\ell)$. The expansion can be summed up to an exponential and is given by
\begin{align}
\label{ansatz}
\psi_\pm(x,\ell)=\psi_\pm(x)\exp\Big[\sum_q\frac{\varphi_q^\pm(\ell)}{\sqrt{n_q}}\phi_q(x)\Big]\quad,
\end{align}
where we defined $\phi_q(x)\equiv e^{iqx}b_q-e^{-iqx}b_q\dag$. Notice that $[\psi_\pm(x),\phi_q(x)]=0$.
The flow equation for $\varphi_q^\pm(\ell)$ reads
\begin{align}
\partial_\ell\varphi_q^\pm(\ell)=-\eta_q(\ell)(\Theta(\mp q)+\varphi_{-q}^\pm(\ell))
\end{align}
where $\Theta(x)$ denotes the Heaviside step function. One obtains the following solution
\begin{align}
\label{AppSolution}
\varphi_q^\pm(\ell)=\Theta(\pm q)(c_q(\ell)-1)-\Theta(\mp q)s_q(\ell)\quad,
\end{align} 
with $c_q(\ell)$ and $s_q(\ell)$ defined below Eq. (\ref{TLSolution}). Notice that no Klein factor is needed in the above representation, Eq. (\ref{ansatz}).

We now define normal ordering $\;:\!...\!:\;$ such that the bosonic operators $b_q\dag$ ($b_q$) stand left (right) from the fermionic operators $\psi_\pm^{(\dagger)}(x)$. With the Baker-Hausdorff formula $e^{A+B}=e^Ae^Be^{-[A,B]/2}$ and the relation
\begin{align}
\psi_\pm(x)f(b_q\dag)&=f(b_q\dag+\Theta(\pm q)n_q^{-1/2}e^{iqx})\psi_\pm(x)\quad,
\end{align}
which holds for an arbitrary function $f$, we obtain
\begin{align}
\label{NormalOrderedTransField}
\psi_\pm(x,\ell)=\;:\!\psi_\pm(x,\ell)\!:\;\exp\Big[-\sum_{q>0}\frac{s_q^2(\ell)}{n_q}\Big]\quad.
\end{align}
With this normal ordering procedure, we can thus extract the anomalous scaling factor $L^{-s^2(\ell)}$, which follows in the limit $L\to\infty$ from a constant potential with finite cutoff $q_c$, i.e., $s(\ell)=s_q(\ell)$ for $q\leq q_c$ and zero otherwise. From the scaling behavior of the transformed field $\psi_\pm(x,\ell=\infty)$ - known from bosonization - we infer that the normal ordered field operator scales like a free fermionic field. This will enable us to substitute the {\it normal ordered} transformed field operator by the bare field for weak electron-electron interactions.

To evaluate the Green's function, we need to normal order the operator 
\begin{align}
\mathcal{G}_\pm^<(x,x';t)\equiv\psi_\pm\dag(x',0,\ell=\infty)\psi_\pm(x,t,\ell=\infty),
\end{align}
where the time dependence is defined in the Heisenberg picture with the fixed point Hamiltonian $H^*=\sum_q \tilde\omega_qb_q\dag b_q$. We obtain
\begin{align}
\mathcal{G}_\pm^<(x,x';t)=\;:\!\mathcal{G}_\pm^<(x,x';t)\!:\;g(x-x',t)\quad,
\end{align}
with ($s_q\equiv s_q(\ell=\infty)$)
\begin{align}
g(x,t)\equiv\exp\Big[-\sum_{q>0}\frac{2s_q^2}{n_q}(1-\cos(qx)e^{i\tilde\omega_qt})\Big].
\end{align}

The Green's function $G_\pm^<(x,x',t)\equiv-i\langle\mathcal{G}_\pm^<(x,x';t)\rangle$ thus reads
\begin{align}
G_\pm^<(x,x',t)=-i\langle\psi_\pm\dag(x',0)\psi_\pm(x,t)\rangle g(x-x',t)
\end{align}
where $\langle...\rangle$ denotes the ground-state average of $H^*=\sum_q \tilde\omega_qb_q\dag b_q$.

Up to now, we have assumed a finite interaction cutoff $q_c$ which yields well-defined expressions. For constant interaction potential $v_q=v\Theta(q_c-|q|)$ and $q_c\to\infty$, the remaining expectation value can be evaluated using the Kronig relation which relates $H^*$ to a free fermionic Hamiltonian with linear dispersion.\cite{FootN} With $\tilde\omega_q=v_c|q|$, $v_c$ denoting the charge velocity, and $c_{k,\pm}(t)=c_{k,\pm}e^{\mp iv_c(k\mp k_F)t}$, we obtain for $L\to\infty$ 
\begin{align}
\langle\psi_\pm\dag(x',0)\psi_\pm(x,t)\rangle
=\frac{-i}{2\pi}\frac{e^{\pm ik_F(x-x')}}{\pm(x-x')-v_ct-i0}.
\end{align}
This yields the well-known result for the Green's function of the Tomonaga-Luttinger model.\\

To complete the discussion on the observable flow, we also transform the ladder operators $c_{k,\pm}$ according to $\partial_\ell c_{k,\pm}(\ell)=[\eta(\ell),c_{k,\pm}(\ell)]$. Again, an infinite hierarchy of operators is generated which is only characterized by one $q$-dependent function of $\ell$, $\varphi_q^\pm(\ell)$. Defining $\phi_{k',q}^k=b_q\delta_{k',k-q}-b_q\dag\delta_{k',k+q}$, the series is given by 
\begin{align}
c_{k,\pm}(\ell)=\sum_{n=0}^\infty\frac{1}{n!}\Big[\prod_{i=1}^n\sum_{k_i,q_i}&\frac{\varphi_{q_i}^\pm(\ell)}{\sqrt{n_{q_i}}}\phi_{k_i,q_i}^{k_{i-1}}\Big]c_{k_n,\pm}
\label{observablefix}
\end{align}
with $k_0=k$. The $q$-dependent function of $\ell$ is again given by Eq. (\ref{AppSolution}). Notice that $\psi_\pm(\ell)=L^{-1/2}\sum_ke^{ikx}c_{k,\pm}(\ell)$. For a step function potential $v_q=v\Theta(q_c-q)$ in the limit $L\to\infty$, we therefore have 
\begin{align}
\label{cNormal}
c_{k,\pm}(\ell)=\;:\!c_{k,\pm}(\ell)\!:\;L^{-s^2(\ell)}\quad,
\end{align}
where $s(\ell)$ as defined below Eq. (\ref{NormalOrderedTransField}). 

We will now add an arbitrary external potential to the initial Hamiltonian, Eq. (\ref{TLmodel}). The additional contribution $H_e$ shall consist of a forward and a backward scattering part, i.e., $H_e=H_F+H_B$. 
The forward scattering contribution can be expressed by the bosonic density fluctuations,\cite{FootN} 
\begin{align}
H_F=\sum_qn_q^{1/2}L^{-1}\int dxU^{F,0}(x)(e^{iqx}b_q+e^{-iqx}b_q\dag).
\end{align}
The backward scattering contribution reads 
\begin{align}
H_B=\int dxU_B^0(x)(\psi_+\dag(x)\psi_-(x)+h.c.),
\end{align}  
with the right- and left-moving field, $\psi_\pm(x)$.   

The flow equations for $H_F$ close and the differential equations for the forward scattering potential  read
\begin{align}
\partial_\ell U_q^F(x,\ell)=\eta_q (\ell)U_{-q}^F(x,\ell)\quad.
\end{align}
We thus have $U_q^F(x,\ell=\infty)=\sqrt{K_q}U^{F,0}(x)$ such that the forward scattering amplitude is enhanced for attractive and reduced for repulsive electron-electron interaction.

The flow of the backward scattering operator
\begin{align}
\mathcal{O}_{\pm 2k_F}(x,\ell)\equiv \psi_\pm\dag(x,\ell)\psi_\mp(x,\ell)
\end{align}
is determined by the flow equation of the field operator $\psi_\pm(x,\ell)$, outlined above. 
Defining the $\ell$-dependent Luttinger liquid parameter $K_q(\ell)\equiv \exp(2E_q(\ell))$, we have  
\begin{align}
&\mathcal{O}_{\pm 2k_F}(x,\ell)=\mathcal{O}_{\pm 2k_F}(x)\\
&\times\exp\Big[\pm\sum_q\frac{\text{sgn}(q)}{\sqrt{n_q}}(\sqrt{K_q(\ell)}-1)\phi_q(x)\Big]\notag
\end{align} 
with $\phi_q(x)\equiv e^{iqx}b_q-e^{-iqx}b_q\dag$. Normal ordering the above expression such that $b_q\dag$ ($b_q$) stands left (right) from $\mathcal{O}_{\pm 2k_F}(x,\ell=0)$ yields
\begin{align}
\mathcal{O}_{\pm 2k_F}(x,\ell)&=\;:\!\mathcal{O}_{\pm 2k_F}(x,\ell)\!:\\
&\times\exp\Big[\sum_{q>0}\frac{1-K_q(\ell)}{n_q}\Big]\notag.
\end{align}

Above, it is shown that the normal ordered transformed field operator scales like a free fermion field. For small electron-electron interaction $\sqrt{K_q(\ell)}-1\to0$, the normal ordered backscattering operator can thus be approximated by the bare backscattering operator, i.e., $:\!\mathcal{O}_{\pm 2k_F}(x,\ell)\!:\;\approx\mathcal{O}_{\pm 2k_F}(x)$.

This approximation provides a suitable truncation scheme for the Hamiltonian flow. We can then define an effective $\ell$-dependent backscattering potential as 
\begin{align}
\label{lambda_Back}
U_B(x,\ell)=U_B(x)\exp\Big[\sum_{q>0}\frac{1-K_q(\ell)}{n_q}\Big]\quad.
\end{align}

It is interesting to analyze the asymptotic behavior of $U_B(x,\ell)$ for $L\to\infty$ and $\ell\to\infty$. For this, we will choose a step function potential $v_q=v\Theta(q_c-q)$ with finite interaction cutoff $q_c$ and also drop the spatial component in the following. Considering first the limit $L\to\infty$, the finite interaction cutoff $q_c=2\pi n_c/L$ shall remain constant, such that $n_c$ is proportional to the system size $L$. With $\sum_{n=1}^{n_c}1/n\to \ln n_c +C$ for $n_c\to\infty$, where $C$ is Euler's constant, we readily find ($K=K_q$ for $q\leq q_c$)
\begin{align}
\label{Asymtotics_L}
U_B\propto L^{1-K}\quad. 
\end{align}
Notice that the renormalization of the external potential depends linearly on the interaction strength $v$.

For an infinite system in the limit $\ell\to\infty$, the sum in the exponent of Eq. (\ref{lambda_Back}) turns into an integral according to $\sum_q\to\frac{L}{2\pi}\int dq$. For a constant electron-electron interaction with sharp cutoff $q_c$, we then obtain
\begin{align}
\label{Asymtotics_ell}
U_B\propto \ell^{(1-K)/2}\quad. 
\end{align}

Notice that the asymptotic limit $\ell\to\infty$ and the thermodynamic limit $L\to\infty$ interchange. The commutation of the two limits does usually not hold even if one considers impurity properties. But in contrary to the treatment of, e.g., the spin-boson model of Ref. \onlinecite{Keh96}, here we do not neglect finite-size corrections in the flow equations.

For repulsive electron-electron interaction, an initial delta-impurity strength thus scales to infinity for $L\to\infty$ and $T\to0$. The ground-state of the fixed point Hamiltonian is then given by the ground-state of an electron gas with fixed or open boundary condition.\cite{Sta03} Concerning the operator flow, this change in boundary conditions can be accounted for by the substitution $b_{-|q|}\to b_{|q|}$ in the operator $\phi_q(x)$, defined below Eq. (\ref{ansatz}), i.e., there are only right-movers. With the same normal ordering procedure as outlined for the case of periodic boundary conditions, this yields the well-known Green's function of the Tomonaga-Luttinger model with open boundaries.\cite{Fab95} We thus obtain the exact power law behavior for the spectral weight at the impurity site.\\

So far, we have mapped the TL model with external potential onto a non-interacting bosonic bath with dispersion relation $\tilde\omega_q=v_Fq\sqrt{1+v_q/\pi v_F}$ and a renormalized external potential. The mapping holds in the limit of weak electron-electron interaction. Linearizing the dispersion relation around $q=0$, one can map the non-interacting bosonic bath onto a non-interacting Fermi gas.\cite{FootN} Notice that, up to terms linear in $v_q$ and approximating $\;:\!c_{k,\pm}(\ell)\!:\;\approx c_{k,\pm}$, these are the same fermions we started with, see Eq. (\ref{cNormal}).\cite{FootBulk} 

The conductance $G$ is now given by the Landauer formula for a one-channel, spinless electron system at $T=0$, $G=\frac{e^2}{h}\mathcal{T}$, where $\mathcal{T}$ denotes the transmission probability through the renormalized external potential.\cite{Lan70,Eco81} By this, we assume that the coupling of the interacting system to the non-interacting leads does not contribute to the transmission probability. $\mathcal{T}$ is thus entirely determined by the renormalized potential which corresponds to perfect coupling.\cite{Mas95} Nonadiabatical coupling can be discussed by considering different boundary conditions of the fermionic field operator.

One can also obtain results at finite temperatures if one identifies the asymptotic flow parameter $\ell$ with the inverse temperature squared, i.e., 
\begin{align}
\label{lT}
\sqrt\ell\propto 1/T\quad.
\end{align}
For this, it is crucial to choose the generator $\eta$ ``canonically'', i.e., based on the commutator of a diagonal and a non-diagonal part. Then, in the case of a simple model, the flow parameter $\ell$ is related to the inverse squared energy uncertainty of the basis states of the fixed point Hamiltonian.\cite{Weg94} This interpretation is confirmed in case of the spin-boson model as the effective tunnel-matrix element $\Delta^*$ is asymptotically reached according to $\Delta(\ell)-\Delta^*\propto1/\sqrt{\ell}$.\cite{Keh96} Furthermore, it turns out that spectral functions, initially depending on the energy $\omega$ and on the flow parameter $\ell$, can be characterized by functions of the scale-independent variable $y=\omega\sqrt{\ell}$ for $\ell\to\infty$.\cite{Len96} This also holds for the Tomonaga-Luttinger model, see Eq. (\ref{Connection}). 

Physical quantities should not depend on energy shifts of the spectrum which are smaller than the energy scale given by the finite temperature. Therefore, the flow equations do not have to be integrated up to $\ell=\infty$, but it suffices to halt the integration at a finite $\ell$. Since the electron-electron interaction is exponentially small for large $\ell$ and not marginally relevant at finite temperatures, we can neglect it. This leaves us with a one-particle problem with $\ell$-dependent parameters. The temperature is introduced by substituting $\ell$ by $T^{-2}$. 

The above procedure yields the same result for the conductance as obtained by Ref. \onlinecite{Yue94} in case of a delta impurity, but is also applicable for general potentials. From Eqs. (\ref{Asymtotics_L}), (\ref{Asymtotics_ell}), and (\ref{lT}), we can also infer the finite-size scaling relation $L\propto 1/T$.

Finally, we want to mention that Refs. \onlinecite{Yue94} and \onlinecite{Med02} obtain an effective non-interacting model with an oscillating and bounded effective potential. But whereas these approaches depart from a non-interacting system with an impurity and then include the electron-electron interaction, we start from an interacting system and add the impurity later, which might explain the different fixed-point Hamiltonians.\cite{FootFr} 

To conclude, we have used Wegner's flow equation method to map the Tomonaga-Luttinger Hamiltonian with an external potential onto a non-interacting Fermi gas with an renormalized external potential. The mapping is valid for weak electron-electron interaction and is based on an alternative bosonization formula for the transformed field operator derived by the flow equation method. Calculating the transmission amplitude from the non-interacting representation, the Landauer formula then provides an analytic expression for the one-dimensional two-terminal conductance for finite size systems. Interpreting the flow parameter in the asymptotic regime as inverse temperature squared yieldes an analytic expression of the conductance also at finite temperatures. 
The formalism applies to arbitrary external potential.

The author wishes to thank V. Meden, A. Mielke, and F. Wegner for helpful discussions.


\begin{thebibliography}{99}
\bibitem{Lut74}
A. Luther and I. Peschel, Phys. Rev. Lett. \textbf{32}, 992 (1974).
\bibitem{Mat74}
D.C. Mattis, J. Math. Phys. \textbf{15}, 609 (1974). 
\bibitem{Ape82}
W. Apel and T.M. Rice, Phys. Rev. B \textbf{26}, 7063 (1982). 
\bibitem{Kan92}
C.L. Kane and M.P.A. Fisher, Phys. Rev. Lett. \textbf{68}, 1220 (1992). 
\bibitem{Boc99}
M. Bockrath, D.H. Cobden, J. Lu, A.G. Rinzler, R.E. Smalley, L. Balents, and P.L. McEuen, Nature \textbf{397}, 598 (1999).
\bibitem{Rot00}
M. Rother, W. Wegscheider, R.A. Deutschmann, M. Bichler, and G. Abstreiter, Physica E \textbf{6}, 551 (2000).
\bibitem{Hal81}
F.D.M. Haldane, J. Phys. C \textbf{14}, 2585 (1981).
\bibitem{Moo93}
K. Moon, H. Yi, C.L. Kane, S.M. Girvin, and M.P.A. Fisher, Phys. Rev. Lett. \textbf{71}, 4381 (1993); R. Egger and H. Grabert, {\it ibid.}  \textbf{75}, 3505 (1995).
\bibitem{Fen95}
P. Fendley, A.W.W. Ludwig, and H. Saleur, Phys. Rev. Lett. \textbf{74}, 3005 (1995).
\bibitem{Sol79}
J. S\'olyom, Adv. Phys. \textbf{28}, 201 (1979).
\bibitem{Yue94}
D. Yue, L.I. Glazman, and K.A. Matveev, Phys. Rev. B \textbf{49}, 1966 (1994).
\bibitem{Sen02}
S. Lal, S. Rao, and D. Sen, Phys. Rev. B \textbf{66}, 165327 (2002). 
\bibitem{Med02}
V. Meden, W. Metzner, U. Schollw\"ock, and K. Sch\"onhammer, J. Low Temp. Phys. \textbf{126}, 1147 (2002).
\bibitem{Weg94}
F. Wegner, Ann.~Phys. (Leipzig) \textbf{3}, 77 (1994); F.~Wegner, Physics Reports \textbf{348}, 77 (2001).
\bibitem{Sta03}
T. Stauber, Phys. Rev. B \textbf{67}, 205107 (2003).
\bibitem{Lan70}
R. Landauer, IBM J. Res. Dev. \textbf{1}, 233 (1957); Philos. Mag. \textbf{21}, 863 (1970). 
\bibitem{Tom50}
S. Tomonaga, Progr. Theor. Phys. \textbf{5}, 544 (1950); J.~M.~Luttinger, J.~Math.~Phys. \textbf{4}, 1154 (1963).
\bibitem{FootG}
Notice that $\eta$ does not contain contributions coming from the external potential which is the case in Ref. \onlinecite{Sta03}.
\bibitem{Mat65}
D.C. Mattis and E.H. Lieb, J. Math. Phys. \textbf{6}, 304 (1965).
\bibitem{FootN}
We neglect terms involving the $q=0$-mode, valid for large systems.
\bibitem{Keh96}
S.K. Kehrein and A. Mielke, Ann.~Phys. (Leipzig) \textbf{6}, 90 (1997).
\bibitem{Fab95}
 M.~Fabrizio and A.O. Gogolin, Phys.~Rev.~B \textbf{51}, 17827 (1995).
\bibitem{FootBulk}
This approximation cannot account for bulk Luttinger liquid behavior and it is crucial that the external potential is renormalized linearly in the interaction strength.
\bibitem{Eco81}
E.N. Economou and C.M. Soukoulis, Phys. Rev. Lett. \textbf{46}, 618 (1981); D.S. Fisher and P.A. Lee, Phys. Rev. B \textbf{23}, 6851 (1981).
\bibitem{Mas95}
D.L. Maslov and M. Stone, Phys. Rev. B \textbf{52}, R5539 (1995); V.V. Ponomarenko, {\it ibid.} \textbf{52}, R8666 (1995); I. Safi and H.J. Schulz, {\it ibid.} \textbf{52}, R17040 (1995).
\bibitem{Len96}
P. Lenz and F. Wegner, Nucl. Phys. B \textbf{482} [FS], 693 (1996); T. Stauber, Phys. Rev. B \textbf{68}, 125102 (2003). 
\bibitem{FootFr}
The Friedel oscillations show up in correlation functions after transforming the observables as well, but do not have to appear in the fixed-point Hamiltonian which can - in principal - be chosen freely.
\end{thebibliography}
\end{document}